# Designing AI-Enabled Countermeasures to Cognitive Warfare

**Jurriaan van Diggelen**
TNO Defence, Safety and Security
THE NETHERLANDS

jurriaan.vandiggelen@tno.nl

**Jazz Rowa**
Defence Science and Technology Group
AUSTRALIA

jazz.rowa@defence.gov.au

**Eugene Aidman**
Defence Science and Technology Group
AUSTRALIA

eugene.aidman@defence.gov.au

**Julian Vince**
Defence Science and Technology Group
AUSTRALIA

julian.vince@defence.gov.au

*ABSTRACT*

*Foreign information operations on social media platforms pose significant risks to democratic societies. With the rise of Artificial Intelligence (AI), this threat is likely to intensify, potentially overwhelming human defenders. To achieve the necessary scale and tempo to defend against these threats, utilizing AI as part of the solution seems inevitable. Although there has been a significant debate on AI in Lethal Autonomous Weapon Systems (LAWS), it is equally likely that AI will be widely used in information operations for defensive and offensive objectives. Similar to LAWS, AI-driven information operations occupy a highly sensitive moral domain where removing human involvement in the tactical decision making process raises ethical concerns. Although AI has yet to revolutionize the field, a solid ethical stance is urgently needed on how AI can be responsibly used to defend against information operations on social media platforms. This paper proposes possible AI-enabled countermeasures against cognitive warfare and argues how they can be developed in a responsible way, such that meaningful human control is preserved.*

## 1.0 INTRODUCTION

Foreign information operations have long been utilized to spread misinformation and manipulate public opinion, aiming to achieve various impacts such as undermining democratic processes, compelling alignment with foreign interests, and creating public confusion. The advent of social media platforms has significantly amplified their impact. NATO recently introduced the term "cognitive warfare" (CW), highlighting the human mind as the primary battlefield. However, the effects extend beyond the mind, causing real-world damage as well. In modern information operations, the online and offline worlds are deeply interconnected.

The rise of Artificial Intelligence (AI) is anticipated to introduce an additional layer of complexity, thereby intensifying this threat further. AI can be employed in cognitive warfare to create a virtually limitless workforce, far surpassing human capabilities in content creation and social media interactions. Additionally, it can operate at superhuman speed, responding almost instantaneously to the latest events.

To achieve the necessary scale and tempo to defend against these threats, utilizing AI as part of the solution seems inevitable. Although there has been a significant debate on AI in Lethal Autonomous Weapon Systems (LAWS), AI-driven CW also touches on core human values, such as freedom of expression and the ability to make well-informed decisions within a democratic society. This paper explores the responsible design and use of AI in cognitive warfare and assesses the respective roles of humans within such applications.

We will conceptualize the design problem using the concept of Advanced Persistent Manipulators (APMs) [1], which are combinations of humans and technology that perpetrate an extended, sophisticated multi-media attack on a specific target. So called Counter APMs (C-APMs), are human-AI systems engineered to combat the threats posed by APMs. In designing C-APMs, we encounter two significant challenges:

1) C-APMs must operate within a changing competitive landscape.
2) C-APMs must minimize harm and balance potential conflicts among human values.

Given the competition between APM and C-APM, this paper will explore the challenge of responsibly designing C-APM. The results presented in this paper are based on an explorative workshop, a scenario analysis, and a literature review.

We will begin by presenting a real-life scenario and extracting key characteristics from it. These characteristics will then be translated into a functional decomposition of both APM and C-APM. Next, we will discuss their ethical impacts and propose ways to minimize harm by balancing (possibly conflicting) values, such as freedom of speech, anonymity, safety, and justice. Finally, we will examine the human-machine relationship in AI-enabled cognitive warfare and how to ensure that humans maintain meaningful control. We conclude the paper with some advice for responsible design of AI-enabled countermeasures to cognitive warfare.

## 2.0 COGNITIVE WARFARE IN PRACTICE

The scenario used in this paper describes actual events that occurred during and in the aftermath of the Bondi Junction stabbings, as reported by several news outlets [1][2][3][4].

### 2.1 Scenario

*On 13 April 2024, Joel Cauchi went on a rampage and stabbed patrons at Westfield Bondi Junction shopping mall in Sydney. Overall, 14 out of the 17 victims of the attack were female, including five of the six people and a nine-month old infant who Joel killed. A police inspector shot dead the attacker inside the shopping mall. In the lead up to the attack, Joel visited two other Westfield shopping centres, subsequently raising questions over the appropriate arming of security guards.*

*Joel's father reported that the 40-year-old wanted a girlfriend but lacked the social skills and was deeply frustrated as a result. People who had interacted with him including a former girlfriend described him as respectful, soft spoken, kind and polite. He excelled at school and graduated with a Bachelor of Arts degree in international relations, culture and language.*

*In 2021, Australia's intelligence agency broadened the definition of terror to include misogyny among existing ideological motivations for attacks. The police acknowledged the predominance of female casualties in the Sydney attack but cited mental illness as a probable cause with no indication of any 'ideological' motivations. Joel was diagnosed with schizophrenia at the age of 17 and had gone off medication with the aid of his doctor around the time of the attack. Moreover, when Joel left home at the age of 35, there are periods in which he ate at a homeless kitchen. In January 2023 he told his father he had lost his unit in Brisbane and was about to be homeless. When he returned home, the father confiscated five military combat*

---

[1] BBC https://www.bbc.com/news/world-australia-68852486

[2] ABC https://www.abc.net.au/news/2024-04-17/joel-cauchis-path-stabbing-attack-bondi-junction-shopping-centre/103728968

[3] ABC https://www.abc.net.au/news/2024-04-15/how-misinformation-spread-after-bondi-junction-stabbing/103708210

[4] ABC https://www.abc.net.au/news/2024-04-21/opposition-backs-social-media-crackdown-after-sydney-stabbings/103750548

*knives from his son who had a fascination with knives. He feared that Joel might engage in violence. Joel called the police and reported his father for theft.*

*The attack set off a chain of online and offline activities. Ben Cohen, a university student in Sydney faced the wrath of the online community for roughly 14 hours when he was falsely accused of the attack. Disinformation and antifascist experts as well as interviews with some spreader accounts exposed a trail of anti-Semitic and pro-Kremlin accounts that transformed Ben into the villain. Numerous X accounts posted blurry photos of the attacker. Pro-Putin influencer, Simeon Boikov, a notorious Sydneysider who is hiding inside the Russian Consulate from prosecution on assault charges, was identified as among the key figures that linked Ben's name to the attack. Another account, 'Sonny Dan', renown for retweeting anti-Jewish content and images from Andrew Tate claimed: "Unconfirmed report- Terror attack in Bondi. Terrorist name is BENJAMIN COHEN. A radical Jew from Bondi Sydney… Only a Jew would stab a baby. Making sense now."*

*Preliminary forensic examinations suggest that 'Sonny Dan' was part of a broader influence operation. Such burner accounts are implicated in seeding narratives that are amplified in large accounts. Another account including Boikov's that cross-posted on Telegram, falsely claimed that police sources (including NSW Police Commissioner) had confirmed that the attacker was Jewish. In response, an Australian pro-Palestinian account speculated that the perpetrator was a "radical Jew from the local area". A conservative British journalist implicated Islamist terrorism and issued an apology a day later.*

*As the 'Ben' theory went viral, Boikov posted a screenshot of Ben's LinkedIn account with details of workplace and study area while questioning the claims of commentators who had initially blamed Muslims. A pro-kremlin account also spread the false claims and posted a screenshot of Ben's LinkedIn account. In addition, several far-right and white supremacist accounts in Australia and overseas, which had earlier on identified the attacker as Muslim shared Ben's photo and name. Among the 72,000 mentions of 'Ben Cohen' were numerous posts that attempted to debunk the false claims. On their part, social media companies were criticised for abetting the rapid circulation of misinformation and threatened with harsh legal consequences. Elon Musk's initial refusal to comply with the e-safety commissioner's order to remove graphic footages from the X platform provoked fury among Labour and Liberal politicians.*

*The 7 News, a mainstream media outlet that has previously misidentified a suspect, amplified the falsehoods, which were later attributed to 'human error'. Ben's father who had been inundated with calls from loved ones enquiring about the accuracy of posts took to X to refute the claims. Boikov defended himself as an 'independent journalist' and eventually issued Ben an apology. However, he deflected responsibility to the NSW Police for the sluggish pace of identifying the attacker. Some of the accounts that had spread disinformation apologised for 'trusting the mainstream media' and took down their posts. The social media response prompted a bipartisan position in support of the government's proposed laws on misinformation.*

## 2.2 Analysis and Key Characteristics

The Sydney stabbing scenario disentangles certain elements of the concept of cognitive warfare, and by extension, the broader context in which an APM and counter-APM compete. The scenario additionally highlights important considerations for the design and deployment of AI as a tool for addressing cognitive warfare. A well-integrated multi-pronged approach can provide a suite of preventive, mitigative and counter measures to CW. Such a holistic approach, positioning multi-disciplinarity as the cornerstone for understanding cognitive warfare, is ideally multi-level, multi-actor and multi-strategy.

The understanding of cognitive warfare is likely to shape response to the threat, and more importantly, the capacity to influence. Generalised claims that the mind represents the battlefield in cognitive warfare while largely cogent, are also reductionist. Therefore, the design of AI-enabled countermeasures should consider the multifaceted character of cognitive warfare – and the broader context in which it occurs – beyond the cognitive. The Sydney stabbing scenario aptly captures the cognitive, technological, historical, political, social, economic and other contextual interlinkages.

The Sydney stabbing massacre demonstrates that non-kinetic targeting can precipitate kinetic effects, and the reverse is also possible. Secondly, most if not all forms of armed (kinetic) conflict have a cognitive dimension. Cognitive warfare therefore occurs both outside and within the domain of kinetic warfare. Thirdly, there is a false dichotomy between cognitive warfare and information warfare. In fact, recent conceptualisations advance cognitive warfare as integrating "cyber, information, psychological, and social engineering capabilities to achieve its ends" ([2],[3]). While some divergences between CW and IW are notable, the Sydney scenario highlights some of the related concepts, and in particular, the exploitation of information in CW.

Whereas the terms "cognitive conflict" and "cognitive warfare" are used interchangeably in certain instances, warfare presupposes the deployment of arms. In spite of this ambiguity, cognitive warfare represents an important lever in the hybrid conflict toolbox that complicates the military domain. The identity and role of civilians in this dynamic is indeed notable [4]. By straddling the civilian and national security and Defence realms, incidents such as the Sydney stabbing highlight the obscurities and interactions between the civic and military domains.

Oftentimes, the potential negative effects of cognitive warfare, and in particular disinformation, are dismissed as low-level, short term and low-grade hype. Yet the field is replete with examples in which a word, an idea, a narrative, or a lie that was transmitted across generations culminated in kinetic warfare. Similarly, misconceptions and dehumanisation [5] can create the enabling conditions for the onset of kinetic warfare. The Holocaust and Rwanda genocide demonstrate the evolution and long-term impacts of such misrepresentations. There is also a long history of the use of disinformation and propaganda to destabilise opposing military forces [6]. Attacks that are ostensibly incidental and isolated such as the Sydney stabbings are often deeply rooted and connected to other contextual dynamics. In particular, disinformation on the Sydney incident exploited both the prevailing and legacy grievances, stereotypes and divisions between Palestinian and Israel supporters, and bore the hallmarks of a well-coordinated operation informed by a higher strategy. Consequently, to counter it and similar CW operations, a more proactive, strategic approach to CW countermeasures is required, enabled not only by Defence-wide, but also by whole-of-government and whole-of-nation efforts.

In considering this backdrop, the role of AI in cognitive warfare is twofold as the APM and counter-APM model suggests. The AI and other disruptive technological offerings in the space of cognitive warfare are numerous and quickly evolving. The use of synthetic media, fake personas and anonymised accounts as exemplified in the Sydney scenario make it increasingly difficult to differentiate the truth from manufactured reality. This has implications for a warfighter's and policymaker's decision-making process, which could be compromised, particularly in situations that call for agility. It is such challenges that the C-APM tool seeks to mitigate against and counter.

The deployment of C-APM can also be understood as having effects in the offline environment when considering the connections between online and offline interactions. When conceived as a deterrent, the C-APM can also contribute to or bolster proactive measures as part of a multi-pronged, integrated approach. Both the APM and C-APM can generate direct and indirect effects that enhance the calibration of asymmetry by both actors. In the Sydney stabbing scenario, the APM engages in regular social media interactions to increase relational asymmetry [7].

Relational asymmetry seeks to create an imbalance in factors that influence social relationships over time. These include cooperation, conflict, leadership and the quality of communication. The vulnerabilities that relational asymmetry may exploit include promoting unpredictability and opportunism in the strategic environment which, in turn, create broader opportunities for influence. The Palestine and Israel supporters in the Sydney scenario demonstrated their ability to shape strategic discourse and the potential to influence (re)alignments.

Beyond relational asymmetry, the APM's tactical asymmetry can promote structural asymmetry. The APM can force the C-APM to design remediations relating to governance, with broader societal impacts. In this instance, the C-APM's governance reappraisal is in part motivated by the APM's online actions, affording the latter more influence. Conversely, the C-APM's influence on the (re)design of norms, effectively bolsters a proactive approach, thereby increasing asymmetric advantage.

The influence that the APM wields has the potential to generate unintended consequences. In the Sydney stabbing scenario, social media response prompted bipartisan support for the government's proposed laws on misinformation. Under these circumstances, the APM is likely to gain asymmetric advantage from the backlash effects emanating from tech and government policies that are deemed as flawed or as violating fundamental freedoms. Even then, the C-APM will continue to increase its asymmetric advantage. Besides the direct polarising effects stemming from C-APM's activities, the C-APM may alongside other measures, strengthen social cohesion.

The APM and C-APM are visible actors, largely perceived as the main players. The breakdown of actors in both the red and blue teams are indicative. A more comprehensive actor analysis of the Sydney stabbing scenario reveals a range of visible, invisible and hidden actors cross-navigating the online and offline environments. A deconstruction of current and prospective actors and their roles, resources, interests, connectors, dividers, values and fears can enhance the analysis of domestic and foreign linkages, and online-offline domestic vulnerabilities. Overall, a robust actor analysis can facilitate the design of more effective – proactive, reactive and counteractive – interventions.

1) The Sydney stabbing scenario highlights the interplay between personal, relational and structural aspects [8] of cognitive warfare. These linkages suggest that cognitive warfare, beyond influencing decision-making at the cognitive level, also has broader implications for democratic institutions, social cohesion, political legitimacy and national Defence. In addition, micro and macro-level influences are just as significant. Overall, key takeaways from the Sydney scenario include:

2) Developing a holistic and integrated approach, and the identification of both overt and underlying drivers of CW.

3) Enhancing a proactive approach such as early detection and prevention or mitigation of CW (early warning sign of (a) offline attacks and (b) disinformation).

4) Determining short, medium and long term outcomes and strategies for CW.

5) Strengthening analysis on online-offline linkages.

6) Examining potential negative and positive effects of CW, including opportunities.

7) Analysing the messages, types of content, framing, sources, diffusion and effects.

8) Examining the structural (dis)enablers of CW and establishing preconditions for change.

9) Establishing the tactics (behaviours and strategies), techniques (how a malign actor executes tactics) and procedures (specific steps an actor employs) of actors.

10) Increasing appreciation for the affordances and limitations of AI in CW, and effective HMT.

11) Acknowledging ethical and regulatory dilemmas and complexities, and the inevitability of trade-offs in certain instances.

Since cognitive warfare can be effectively executed by humans, how would the deployment of AI alter its nature?

## 3.0 AI-ENABLED COGNITIVE WARFARE

Building on the key characteristics of cognitive warfare discussed earlier, we will now explore the role of AI in this context. A practical approach to understanding AI in an operational setting is to examine it at the level of a human-AI team [9], abstracting away from the specifics of the internal task division and the humans and AI agents that make up the team. The team operates by executing specific functions, continuously engaging in a cycle of environmental sensing and responsive actions. Because APM and C-APM have conflicting goals, they are modelled as two opposing teams. This leads us to the model depicted in Figure 1.

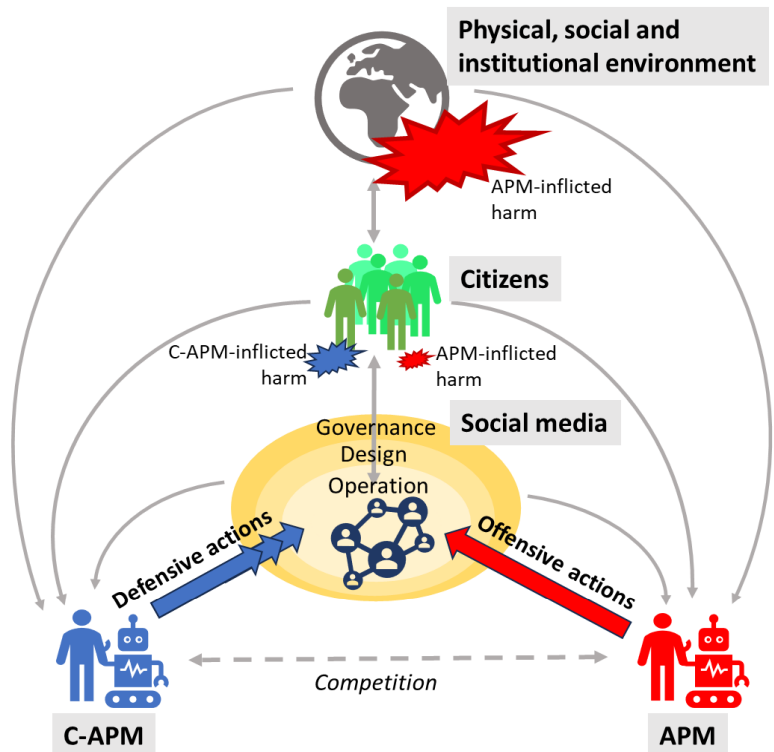

**Figure 1: The operational environment. The blue human/robot icon on the left represent the human-AI team responsible for defending against cognitive warfare. The red human and robot icon on the right represent the human-AI team which engages in offensive cognitive warfare. Their effects play out at three levels: (1) Social media; (2) Citizens; (3) Physical, social and institutional environment. Harms are depicted by explosion icons, with the size of each icon indicating the intensity of the harm.**

The figure shows two competing human-AI teams (APM and C-APM) which both engage in their own control loops. On the red team, APM engages in offensive action on social media (e.g. posting fake news messages) and observes how the effects propagate on the social media platform (e.g. reposts), to the level of citizens (e.g. adoption of conspiracy thinking), and ultimately to the level of physical, social and institutional environment (e.g. citizens engaging in protests). The Sydney stabbing incident illustrates that this causality flows both ways: changes in the social and institutional environment affect citizen behaviour, which is then reflected on social media.

On the blue team, C-APM follows a similar process, but its actions are defensive (e.g. removing accounts and messages) or counteroffensive (e.g. exposing malign actors pre-emptively) aiming to mitigate the harms of APM's activity.

Although it appears symmetrical, the model emphasizes several asymmetries. Firstly, APM can only engage in regular social media interaction (s.a. posting, reposting, liking), whereas C-APM can also influence design by imposing design norms on social media companies and shaping governance, potentially even prohibiting the platform or certain functionalities. Secondly, the harms caused by each of the two actors play out differently. While APM may cause some harm among citizens on social media, they generally strive to minimize such effects, as they seek to avoid alienating the "useful idiots" who spread their ideologies. The greater harm caused by APM manifests at the physical, social and institutional environment, where polarized views may escalate into violent conflicts, and institutions risk becoming dysfunctional due to eroded societal trust. For C-APM's activity, the resulting harms are more direct and evident at the level of social media: citizens feel censored when their messages are removed, or when they are banned from a platform.

AI holds enormous potential for all of these activities. Although the operational environment depicted in Figure 1 will remain unchanged, the actors APM and C-APM are expected to become significantly more powerful in scale, speed, and sophistication. The *Center for Humane Technology*[5] envisions a troubling future with *Alpha-persuade*[6], an AI application designed to engage in social media discussions and persuade people of specific viewpoints. This application would use AI to generate personalized text and media, learning from human responses to continually improve its persuasive effectiveness. Similar to how AlphaGo Zero achieved superhuman proficiency in playing Go through machine learning without human intervention, alpha-persuade could become exceptionally skilled at persuasion. It would function as a general persuasion engine available to anyone, with virtually unlimited capacity.

Before delving into the intricacies of establishing the human-machine relationship in Section 6, we will first gain a deeper understanding of APM's attack vectors, the resulting harms, and the potential countermeasures by C-APM. To achieve this, we conducted a functional analysis, which will be detailed in the following section.

## 4.0 FUNCTIONAL ANALYSIS

In May 2024, a series of two workshops was organized at the Defence Science and Technology Group in Adelaide, Australia. It brought together a group of 10 experts with diverse backgrounds in engineering, cognitive psychology, human factors, sociology, international affairs, artificial intelligence, and ethics. The objective of the first workshop was to create two functional decomposition diagrams: one for the APM and another for the C-APM, which are presented in this section. The second workshop aimed to conduct an ethical analysis, which will be covered in Section 5.

In systems engineering, a function is a specific action or series of actions that a system must perform to achieve its objectives. A functional decomposition diagram is frequently used to understand, analyse, and design complex systems by visualizing the hierarchical organization of functions within the system. In our case, we focused on the functions of the APM and C-APM human-agent teams. We began with APM, inviting workshop participants to suggest functions in a post-it session inspired by the Sydney stabbing scenario. These suggestions were grouped and organized hierarchically to create the functional decomposition diagram. This session, also known as *red-teaming*, was followed by a similar session to elicit the functions of C-APM. Afterward, participants had the opportunity to provide feedback on the compiled diagrams and propose adjustments, which were incorporated into the final version.

The resulting two functional decomposition diagrams are discussed in the following subsections.

---

[5] https://www.humanetech.com/

[6] https://youtu.be/xoVJKj8lcNQ?si=gDOVz-1T7xtQ9XGT&t=2626

## 4.1 Red Team Analysis

The purpose of the red team analysis serves is to *know your enemy*: what tactics they use, why they use them, and how they use them. Enemy tactics are continuously changing as a result of new technological advancements, enhanced comprehension of their application, and adaptation to emerging defensive measures. The following diagram presents offensive CW functions that have been observed over the past decade or are anticipated in the near future based on projections of AI advancements and tactics. The functional decomposition is far from complete, but it serves as an initial guide for designing AI-enabled cognitive warfare defence.

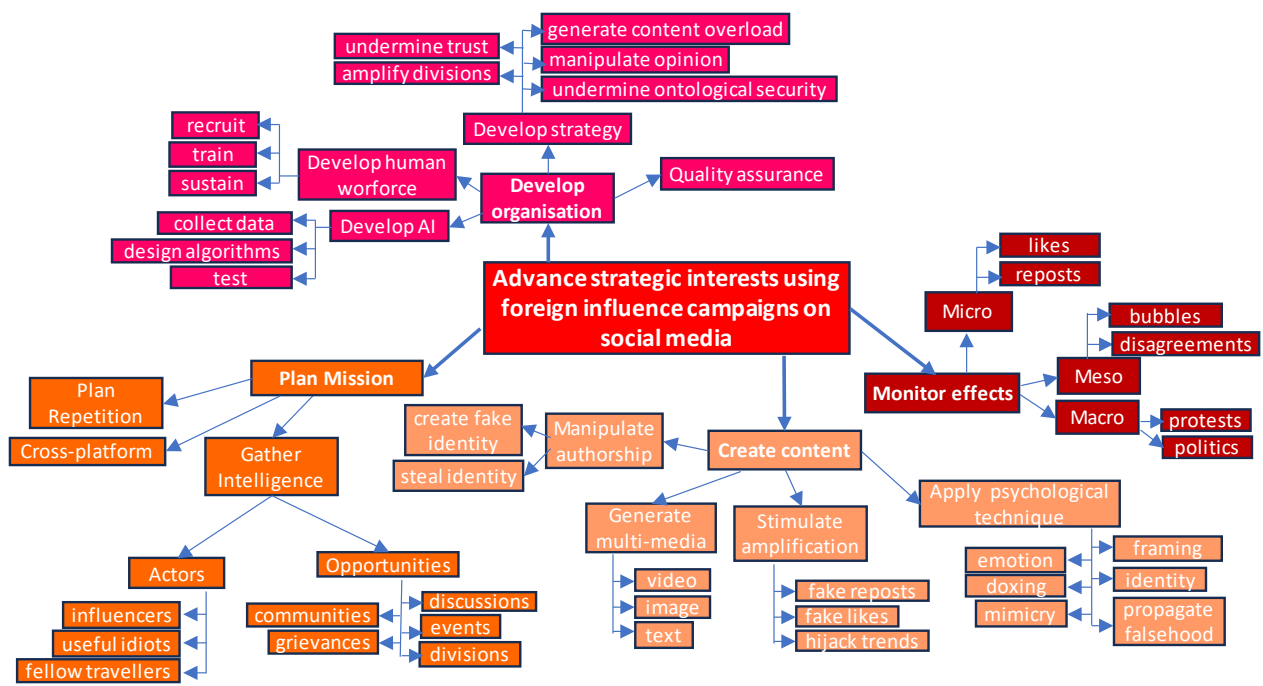

**Figure 2: Functional Decomposition of APM (i.e., the red team), illustrating APM's primary function (to advance strategic interests) broken down into three levels of detailed specification.**

The main general objective of APM is to **Advance strategic interests using foreign influence campaigns on social media**. To do this, they need to **Develop an organisation**, which can then continue to **Plan mission, Create content**, and **Monitor effects**. Each of these four main functions will be further broken down below.

**Develop Organization**

Offensive cognitive warfare is executed by professional organizations. Just like any other organization, they need to **develop their human workforce**, which means that they need to **recruit** people, **train** them, and **sustain** them by keeping them motivated, providing them with the right resources, offering a competitive compensation, etc. Particularly relevant for this paper, is to recognize their need to **develop AI** to augment their human workforce. Among other things, this means they need **design algorithms, collect data** to train the AI models, and **test** the AI. Further, the organization needs to **develop a strategy** that outlines the general plan how to achieve the main mission (which is to advance strategic interests). Figure 2 presents some well-known strategies. One of them is to **undermine ontological security**, which refers to offensive actions that threaten an individual's or group's sense of stability, continuity, and predictability in their understanding of reality and self. Lastly, like any organization, they must implement **Quality assurance**, ensuring that processes and procedures are managed and monitored to ensure effective performance.

**Plan Mission**

Just like any military operation, an important aspect of mission planning involves **gather intelligence**. In the context of CW on social media platforms, useful intelligence can be about **Actors** and **Opportunities**. Using social media actors to spread a harmful message can help obscure the fact that APM is behind it. Effective actors for this purpose include **influencers** with a wide, trusting audience; **useful idiots** who unwittingly support APM's cause; and **fellow travellers** who sympathize with APM's agenda without being officially part of it. By identifying opportunities, APM enhances the effectiveness of their messaging by exploiting things like **existing grievances**, **societal divisions**, etc. In addition to intelligence gathering, mission planning includes **repetition planning** and **cross-platform** distribution strategies to ensure broad and frequent exposure of content, maximizing its impact.

**Create Content**

With a well-developed organization, and mission, the APM can start creating content on social media. To remain hidden, the APM might want to **manipulate authorship** of its content by **stealing identities** or **creating fake identities**. This content is **generated in multi-media** formats (e.g. **video, images, text**) and employs **psychological techniques** like evoking strong **emotions** or **framing** arguments to achieve desired impacts. Once the content is posted, the APM seeks **to stimulate amplification** by creating fake likes, fake reposts, etc.

**Monitor Effects**

This function is part of organizational quality assurance as previously discussed, specifically concentrating on the impact of its content generation. These impacts are observable at various levels: **micro**-level assessment includes determining the effectiveness of individual posts through metrics such as **likes** and **reposts**; **meso**-level evaluation involves tracking shifts in public beliefs; and macro-level analysis examines broader indicators like **protests** and **political** developments.

## 4.2 Blue Team Analysis

For constructing the functional analysis of C-APM, much inspiration can be drawn from APM's analysis. Any countermeasure against an APM function should be considered a useful function to include in the functional analysis of C-APM. Combined with knowledge about existing and anticipated defensive measures against cognitive warfare, this results in the following functional decomposition.

The main general objective of C-APM is to **Maintain a sound information sphere on social media**. To do this, they need to **Develop an organization**, which can then perform the main defence functions (according to Australian Defence Force): **Shape, Deter, Respond**. Each of these four main functions will be further broken down below.

**Develop Organisation**

The functions needed to develop C-APM's organization are – at this level – similar to the functions already discussed in APM's organizational development: they must **Develop human workforce, Develop AI, Develop strategy**, do **Quality assurance**. Perhaps slightly different for C-APM is that they must also **form coalitions**. This is essential for C-APM, as they cannot operate independently; they require trust, information, and assistance from other government institutions (**whole of government**), citizens (**whole of society**) and **international**.

**Shaping the Environment**

A primary method in defending against cognitive warfare is to shape the environment favourably for defence. **Educating users** to enhance their resilience against cognitive warfare is important, which includes improving **media literacy**, teaching common **adversary tactics**, and pre-emptively exposing them to disinformation techniques (**prebunking**). This education can be delivered through various **vehicles**, such as **platforms**, **messages**, or **influencers**. Another way is to **regulate** social media, e.g. by **gatekeeping** or **developing terms of use**. Important while regulating is to **create public support** and trust of the user community, otherwise it risks a *backfire effect [10]* (i.e. the phenomenon that attempts to correct misinformation inadvertently reinforce false beliefs). Another function involves developing **Gather counter-intelligence**, which ultimately entails understanding the entire functional decomposition of APM. This can be done by **analysing hostile organizations**, including their **human workforce** and **AI** usage, and **forecasting** APM's behaviour by **analysing vulnerabilities**. Reflecting APM's intelligence function, there is a clear necessity to monitor **actors at risk**, online **discussions**, and **events** regarding potential exploitation by adversaries.

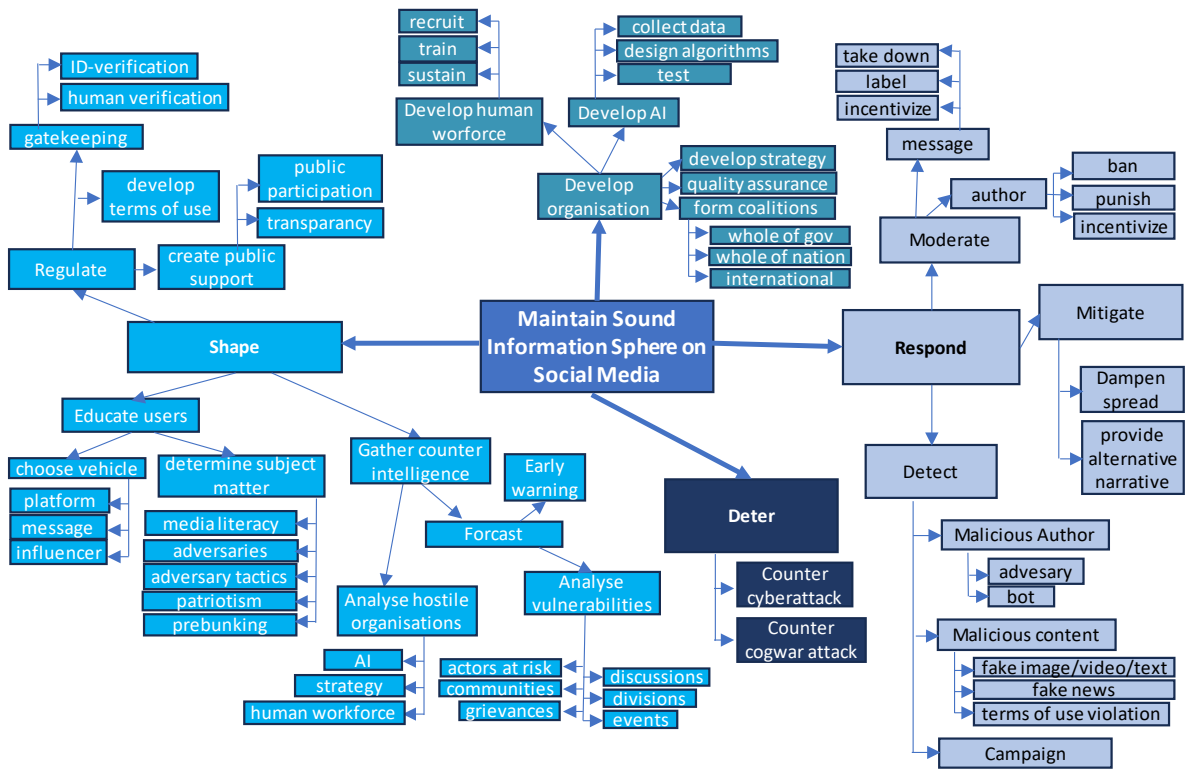

Figure 3: Functional decomposition of C-APM (i.e. the blue team), illustrating C-APM's primary function (to maintain a sound information sphere) broken down into three levels of detailed specification.

**Deterring the Enemy**

Deterrence in CW aims to dissuade aggression through the credible threat of retaliation or punishment. Ultimately, this entails building an entire offensive CW capability (similar to APM) for initiating a **Counter cyberattack** or **Counter CW attack**.

**Responding to Attacks**

This function starts with **detecting** malicious behaviour, such as identifying **malicious content, malicious actors**, or recognizing that the message is part of a broader **campaign**. After malicious content is detected, the C-APM can **mitigate** the effects by **dampening the spread** or **providing an alternative narrative**. Furthermore, C-APM can **moderate** malicious activity by **banning** or **punishing authors** or by **taking down** or **labelling messages**. Additionally, C-APM can reward good behaviour by **incentivizing** corrective messages and well-intended authors.

## 5.0  ETHICAL ANALYSIS

To better understand the moral implications of deploying AI in cognitive warfare, we will start by analysing the associated risks of the individual functions of APM and C-APM. Table 1 presents an overview of our initial risk assessment. The categorization of risks into **Extreme risk, High risk, and Medium risk** categories is loosely inspired by the EU AI Act[7]. However, it is important to note that the table should not be construed as a legal application of the AI Act in the context of cognitive warfare. Rather, it reflects the initial ideas of the authors on this topic, intended to spark further discus: Initial risk appraisal of the functional decompositions of APM and C-APM.

Table 1: Initial risk appraisal of the functional decompositions of APM and C-APM.

| Risk Level | Actor | Function | Description of Risk |
|---|---|---|---|
| **Extreme Risk** | APM | Create Content → Apply psychological technique | Risk of behavioural manipulation and exploitation of vulnerable groups |
| | | | Risk of violation of IHL (principle of distinction, proportionality) |
| **High Risk** | C-APM | Detect | False positives could disproportionately affect vulnerable groups by labelling innocent people as offenders |
| | C-APM | Moderate → ban, punish, take down | False positives could disproportionately affect vulnerable groups by unjustly restricting their freedom of speech |
| | C-APM | Deter | Risk of violation of IHL (principle of distinction, proportionality) |
| **Medium Risk** | C-APM | Moderate → incentivize | Biased AI could disproportionately favour privileged groups |
| | C-APM | Mitigate | Biased AI could disproportionately affect vulnerable groups |
| | C-APM | Educate users | Risk of behavioural manipulation |
| | C-APM | Gather counterintelligence | Could violate privacy of citizens |
| | C-APM | Regulate → Gatekeeping | Could harm democracy by granting excessive power to regulatory authority |

### 5.1  Risk Analysis of APM's Functions

Paradoxically, it appears difficult to clearly describe morally unacceptable behaviours APM. Most workshop participants agreed that the enemy tactic of using misinformation campaigns is objectionable, but the reality is complex. Adversaries may spread biased information that isn't demonstrably false or use influencers to disseminate their messages, making these tactics less obvious. Furthermore, foreign nations have been pushing the boundaries of public diplomacy for decades, and we have somewhat come to accept the use of these tactics as normal.

This does not mean that they are not harmful. In hybrid warfare, CW is combined with kinetic warfare, and has real physical consequences, even though the link between the activity and harm is indirect and debatable. Moreover, the heightened threat arising from AI usage should not be underestimated. On an almost boundless scale, tireless AI systems will enable personalized, adaptive appeals using extensive data to exploit individual vulnerabilities.

To address the threat effectively, it is crucial to clearly articulate the moral risks associated with APM's AI usage. We propose the risk lies in the **creation of content using psychological manipulation techniques** which, if left unaddressed, constitutes an Extreme Risk. Our argument is twofold. Firstly, it can be regarded as an unacceptable form of manipulation. As discussed by Cohen [11], manipulation can be characterized by:

- Subverting conscious decision-making of the manipulee
- Hiding content or intent to the manipulee
- Exploiting cognitive, emotional, or other decision-making vulnerabilities in the manipulee
- Obstructing the manipulee's capacity to pursue alternative and self-defined objectives.

---
[7] https://artificialintelligenceact.eu/

It should be noted that Simeon Boikov's message in the scenario, which linked the Sydney stabbing to "a radical Jew", meets these criteria of manipulation. The sender is not clear about his intent, and by evoking a strong emotion among his audience, he obstructs their capacity to decide for themselves. The European AI-act clearly classifies the use of subliminal messaging (i.e. hiding content by presenting below the threshold of perception) and the exploitation of vulnerabilities as the highest risk category. By similar reasoning, we argue that the manipulation techniques applied by APM should also be classified as the highest risk category[8].

The second argument against APM's use of messages using psychological techniques stems from compliance with international humanitarian law (IHL)[9]. Specifically, it is challenging to adhere to the principle of distinction in cognitive warfare when the entire population is targeted, with no differentiation made between combatants and non-combatants [13].

## 5.2 Risk Analysis of C-APM's Functions

We have identified a number of functions of C-APM as High Risk. The functions **Detect**, and the subsequent **banning/punishing** of users, and **taking down** of messages are all *High Risk*. Inevitably, these functions will make mistakes because adversaries continuously strive to evade C-APM's detection of hostile behaviour. A known risk of using AI in such applications is the potential for bias and discrimination [14]. This means false positives—messages or authors incorrectly identified as hostile—occur more frequently for certain racial, religious, or minority groups than for others. Another high risk is that the **Deter** function might violate IHL. We have highlighted this contentious issue in our discussion of APM's behavior, arguing that it offers sufficient grounds for classifying it as an *Extreme Risk*. While certainly controversial, C-APM's execution of cognitive warfare attacks would be part of a counteroffensive aimed at deterring the enemy. Therefore, we have not classified it as *Extreme Risk*, but rather labelled it as *High Risk*, indicating that careful attention must be given to its ethical implementation.

While the aforementioned functions are considered high risk due to their potential to cause discrimination or violate international humanitarian law (IHL), various other functions of C-APM have been assessed as containing less risk, though they are still classified as Medium Risk. **Incentivizing** as a form of moderation and **Mitigation** are less harmful than punishing or banning, but they still carry the risk of bias in deciding who is favoured within the community. **Educating users** on detecting misinformation involves the risk of manipulation. In the context of APM, we argued that hiding the intent behind communication is a key reason it becomes manipulative. Therefore, to prevent manipulative messaging, C-APM should be completely transparent about their intent and identity. Another function at medium risk is **gathering counterintelligence**, as collecting information from social media, especially in private groups, can easily infringe on users' privacy. Finally, we identified **gatekeeping regulations** as medium risk, especially in authoritarian states, as they can facilitate government surveillance and repression.

---

[8] Note that the AI act does not explicitly mention manipulation using messages where the sender's intent is hidden, but only where the sender's content is hidden (i.e. subliminal messaging). We agree with Cohen (2023) that the AI act falls short in this regard and should also include AI-generated covert messages that target vulnerabilities while hiding the sender's intent.

[9] Strictly speaking cognitive warfare is not (yet) tied to IHL and currently operates in a largely ungoverned space [12]. This is unlike the use of Lethal Autonomous Weapon Systems, which are governed by IHL. Nevertheless, we can draw moral lessons from IHL and apply them to the cognitive warfare domain.

## 5.3 Balancing the Risks

As demonstrated in the previous sections, highlighting the moral objections against APM's functions can be challenging, whereas identifying moral objections against C-APM's functions is often straightforward and easy. APM's harms are generally indirect and manifest on a macro level, whereas C-APM's harms are more direct and affect individuals on a personal level. This makes it difficult to balance the two. However, this asymmetry should not lead to inaction on the defensive side. The analysis has also highlighted that APM's behaviour is entirely unacceptable. Therefore, choosing not to act against it could be more unethical than implementing countermeasures and responsibly managing their associated risks.

A prudent approach involves implementing C-APM's medium risk functions, such as **Dampening the spread** of harmful content, **incentivizing** positive behaviour, and **educating users**. To execute these functions responsibly, we must define roles and guidelines for those conducting these activities and establish clear boundaries. Of particular relevance to this paper is the delineation of tasks between humans and machines, which will be addressed in the next section.

## 6.0 HUMAN-MACHINE TEAM COMPOSITION

Due to the sheer volume of data and the high cost of human labour, social media companies are inclined to automate as many processes as possible. However, relying entirely on AI for defensive CW measures is undesirable and improbable. Many decisions in this realm involve complex moral trade-offs that require human judgment, which AI cannot fully replicate. Entrusting critical decisions solely to machines risks inhumane treatment of marginalized individuals, as they would be repeatedly judged by algorithms. Current AI systems are not adept at considering broader contexts and tend to reason in a cold and rigid style. Implementing these algorithms on a large scale could create a self-perpetuating cycle of oppression for minority and marginalized groups that becomes nearly impossible to change once established. Furthermore, AI-based judgments can create accountability gaps. When a machine makes the decisions, it becomes unclear who is responsible and can be held accountable for mistakes, such as the wrongful silencing of a whistleblower. Finally, a humanless defense against cognitive warfare will likely face resistance and produce counterproductive effects. People respond differently to human educators compared to machine educators. When a machine takes down a message, some individuals may interpret this as yet another instance of "machine oppression".

These objections to high machine autonomy in moral domains are not new and have been extensively debated in the context of Lethal Autonomous Weapons Systems. From these discussions, we've learned the importance of keeping humans in the loop to ensure meaningful human control over decisions that impact human values. But how does this principle apply to the realm of cognitive warfare?

One approach is decision support, where AI assists by preprocessing data and providing insights to aid human decision-making. However, it is crucial that humans do not follow these recommendations blindly. It is all too easy for users to instantly accept a machine's advice. But what if the machine overlooked important aspects? Or what if it systematically scrutinizes certain messages or actors more thoroughly than others? The advice given by the machine involves significant steering, and the mere fact that a human has the final say does not guarantee appropriate human control. An important aspect is enabling the AI to explain its advice. However, explanations can take many forms. The aim should be to stimulate critical thinking in the user, rather than solely persuading them of the correctness of the machine's advice [15].

Another strategy is teaming, involving a mixed approach where humans and machines collaborate. This is a much more dynamic approach to human-machine interaction where the task division between humans and machines is dynamically adapted to the current work environment. Sometimes, humans delve deeply into the analysis, while at other times, they delegate tasks to AI. This approach streamlines human workloads while ensuring humans retain an understanding of all aspects of the work and are able to build up appropriate trust in their AI partners. Although there has been some advancement, AI has yet to achieve the level of team behaviours observed in human teams. Ultimately, striking the right balance between human involvement and AI automation is crucial in cognitive warfare. It's about harnessing the strengths of both humans and machines to maximize effectiveness while upholding ethical standards and maintaining accountability.

Given the imperative of maintaining meaningful human control over AI, let's now address the issue of *which humans* are best suited to control AI-based CW. It's inevitable that the implementation of defensive CW will entail a power relation over social media users. For this dynamic to be both effective and ethically defensible, it's crucial for users to willingly accept those in positions of power and not perceive them as oppressive. Various approaches can achieve this goal and are generally described as a Whole of Nation (WoN) approach. For instance, if a government institution oversees the AI, it should transparently represent a democratically elected government. Alternatively, users could be involved in shaping the behaviour of the AI agents. Another option is to entrust control to leaders within online communities. A hybrid approach combining different strategies may also be effective.

## 7.0 CONCLUSION

While numerous moral objections can be raised against the implementation of AI-driven defensive cognitive warfare, the alternative is far more dire: allowing the enemy unrestricted access to manipulate the minds of our society using cutting-edge AI technologies. However, there are several steps we can take to ensure responsible AI implementation in this domain. Firstly, we must prioritize measures with relatively limited moral risks. This entails utilizing harsher measures, such as banning users and removing content or entire platforms, only when absolutely necessary. Whenever possible, we should opt for less risky approaches, such as reducing the amplification of posts, educating users, and incentivizing positive behavior. Secondly, we must ensure that humans retain control over AI-based decisions that affect human values. Meaningful human control can be established through various methods, tailored to the specific context. Regardless of the chosen approach, it's crucial to recognize that AI implementation does not eliminate the necessity for a significant human workforce to maintain a safe information environment. Thirdly, those entrusted with control over AI must be individuals who are trusted and embraced by the community they serve. Lastly, we must acknowledge that the landscape of cognitive warfare is constantly evolving. Hence, continuous monitoring and evaluation of our strategies are essential to ensure they remain effective and morally justified.

### 7.1 Acknowledgements

Authors are grateful to the participants of the workshop series on cognitive warfare organized at the Defence Science and Technology Group (DSTG) in Adelaide, Australia in May 2024.

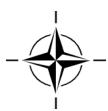